# Automatic control of gold nanoparticle clustering in plasmonic nanopores based on Raman spectroscopic feedback

Iuliia Sheinman[1], Yingqi Zhao[1], Zhendai Huang[2], Kuo Zhan[1], Zuyan Chen[2], Ilkka Nissinen[2], Shuai Li[2,3], Jianan Huang[1,4,*]

1. Faculty of Medicine, University of Oulu, Oulu, 90220, Finland
2. Faculty of Information Technology and Electrical Engineering, University of Oulu, Oulu, 90570, Finland
3. VTT, Technical Research Centre of Finland, Oulu, 90590, Finland
4. Faculty of Biochemistry and Molecular Medicine, University of Oulu, Oulu, 90220, Finland

* Email: jianan.huang@oulu.fi

**Abstract**

Plasmonic nanopores have been used to trap single nanoparticles using electric bias for single-molecule surface enhanced Raman spectroscopic (SM-SERS) detection and sequencing. While they exhibit high sensitivity, removal of nanoparticle clusters in the nanopore is so labor intensive that it hampers biosensing applications. Here, our approach is to use a microcontroller to change the electric voltage to remove the nanoparticle clusters in the nanopore with feedback by the Raman spectra. The SM-SERS spectra of the molecules adsorbed on nanoparticle surface can be acquired over a sufficiently long period, so that the microcontroller can analyze them with <100 ms and trigger the voltage change to remove the nanoparticle cluster. Our system is promising for automatic and high-throughput SM-SERS detection and analysis of DNA and proteins.

## 1 Introduction

Surface-enhanced Raman spectroscopy (SERS) enables the acquisition of molecular vibrational spectra that serve as unique spectral fingerprints for the identification of a wide range of analytes. Under optimized conditions, SERS has demonstrated sensitivity down to the single-molecule level. This combination of ultrahigh sensitivity and chemically specific molecular identification distinguishes SERS from established analytical techniques such as mass spectrometry and nuclear magnetic resonance (NMR) spectroscopy. In contrast to mass spectrometry, SERS does not require analyte ionization and provides direct access to vibrational information. A SERS system capable of detecting various molecules, such as proteins and DNA, on the surface of nanoparticles may be employed in the development of biosensors.

Recently, we have developed a single-molecule SERS (SM-SERS) system by trapping single gold nanoparticle in a gold nanopore to detect DNAs and peptides.[1] The biomolecules were firstly adsorbed on the gold nanoparticle before trapping. The nanoparticle transport is driven by an electric potential difference applied between the Trans and Cis chamber on the chip. Because the surface charge of the nanoparticle is similar to those on the nanopore sidewall, the electrophoretic force on the nanoparticle will be balanced by the electroosmotic force in the nanopore to stop the particle transport in Z direction. Meanwhile, the plasmonic gradient force of the gold nanopore under laser illumination can pull the nanoparticle to the edge of the pore sidewall, which enables trapping of the gold nanoparticle in the nanopore. However, nanoparticle clogging in the nanopore happened when the nanoparticle sticking on the nanopore wall, which need manual reversing the electric bias to remove the clogged nanoparticles.

Here, we describe a system that enables automatic changing of the applied bias voltage to remove the particle clog based on the acquired SM-SERS spectra. By directly reading the SM-SERS spectra from CCD and real-time analysis, we demonstrated that the nanoparticles producing a signal that is not characteristic of a single molecule can be removed from the nanopore. Our system enables automatic and high-throughput SM-SERS detection and analysis of DNA and proteins.

## 2 System Composition

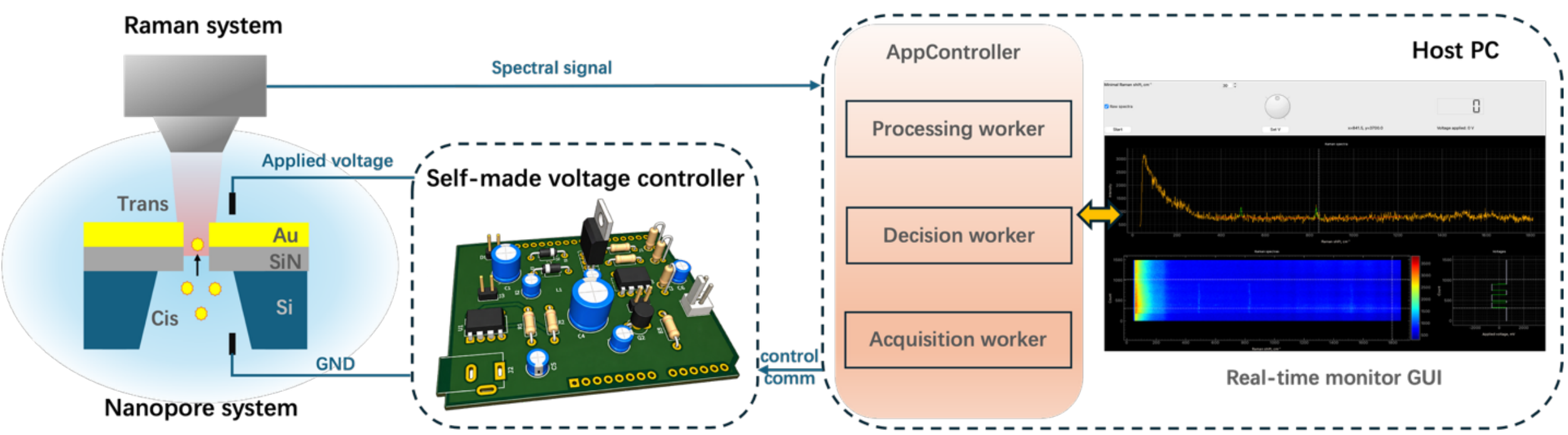

*Figure 1. Architecture of the real-time Raman-feedback voltage-control platform.*

## 2.1 System Overview

Figure 1 presents the overall architecture of the real-time Raman-feedback voltage-control platform. The platform integrates a nanopore-based electro-plasmonic trapping system with Raman spectroscopy and closed-loop voltage control to manipulate SERS signals associated with gold nanoparticles (AuNPs). It comprises two interconnected components: a nanopore-SERS experimental setup for generating and measuring Raman signals, and a software-hardware control system for synchronized spectral acquisition, voltage recording, real-time spectral analysis, and feedback control.

**Materials**

Non-functionalized gold nanoparticles (AuNPs) with an average diameter of 50 nm were purchased from Sigma-Aldrich (753645-25ML; particle concentration: $3.5 \times 10^{10}$ particles/mL). A SYLGARD™ 184 Silicone Elastomer Kit was used to fabricate polydimethylsiloxane (PDMS) microfluidic channels. Silicon wafers containing 100-nm-thick SiN membranes were purchased from MicroChemicals GmbH.

**Nanopore Device Fabrication**

Gold nanoholes were fabricated on low-stress SiN membranes supported by silicon substrates. The SiN membrane window had dimensions of $1 \times 1$ mm² and a thickness of 100 nm, while the supporting silicon chips had dimensions of $1 \times 1$ cm². A 2-nm-thick titanium adhesion layer followed by a 100-nm-thick gold layer was sputter-deposited onto the front side of the SiN membrane, while a 20-nm-thick gold layer was deposited onto the back side.

Nanopores with a diameter of approximately 200 nm were subsequently milled from the back side of the membrane using focused ion beam (FIB) milling (FEI Helios DualBeam). The size and morphology of the fabricated nanopores were characterized from the front side using scanning electron microscopy (SEM). Finally, the nanopore chips were integrated into a custom-made PDMS microfluidic chamber for subsequent experiments.

**Solution preparation**

The AuNP suspension used in the experiments was prepared by combining 300 µL of the 50-nm AuNP stock suspension with 800 µL of 5 % PBS in a microcentrifuge tube. The prepared suspension was stored under refrigeration until use. During the measurements, the AuNP suspension was introduced into the cis compartment, whereas the trans compartment was filled with 5 % PBS without AuNPs.

**Raman measurement system**

The nanopore-SERS chip was mounted under a Raman microscope. The chip comprised a silicon substrate with a central thin SiN membrane and was sealed within a PDMS structure to form two fluidic compartments, referred to as the cis and trans chambers (Figure 1). The cis chamber contained the prepared 50-nm AuNP suspension, while the trans chamber contained 5 % PBS without nanoparticles.

Ag/AgCl electrodes connected to the bias-voltage source were inserted into the two chambers to apply a potential difference across the nanopore. When an AuNP approached the nanohole sidewall under laser excitation, plasmonic coupling between the AuNP and the nanohole generated a confined electromagnetic hot spot, enabling electro-plasmonic trapping and SERS detection[1]. The applied bias voltage was adjusted to remove unsuitable particles or hold suitable AuNPs near the nanohole sidewall. Raman spectra were recorded by the Raman camera (Andor Technology, Newton series Electron Multiplying Charge-Coupled Device, DU970P_BVF) and analyzed in real time using a custom Python application.

**Raman-feedback voltage-control platform**

During operation, Raman spectra are continuously acquired from the nanopore region and transferred to a Python-based control program running on the host PC. Each newly acquired spectrum is preprocessed and analyzed to extract peak and baseline features, which are subsequently used to determine the current operational spectral state. The decision module evaluates the detected state and determines whether the applied voltage should be maintained, reversed, or further adjusted. The resulting command is transmitted to the programmable voltage controller, while the Raman spectra, spectral states, and applied voltage are recorded synchronously and displayed through the graphical user interface.

## 2.2 Self-made Voltage Controller

A self-made voltage controller was developed to provide programmable bipolar bias for the nanopore-SERS chip. The controller consists of a Nucleo-G431RB development board, and a PCB mounted on top of the board (see Figure 2(a)). The PCB converts the microcontroller DAC output into a bipolar analog voltage, which is applied to the nanopore device through an output connector. The output voltage ranges from -3 to +3 V in increments of 100 mV.

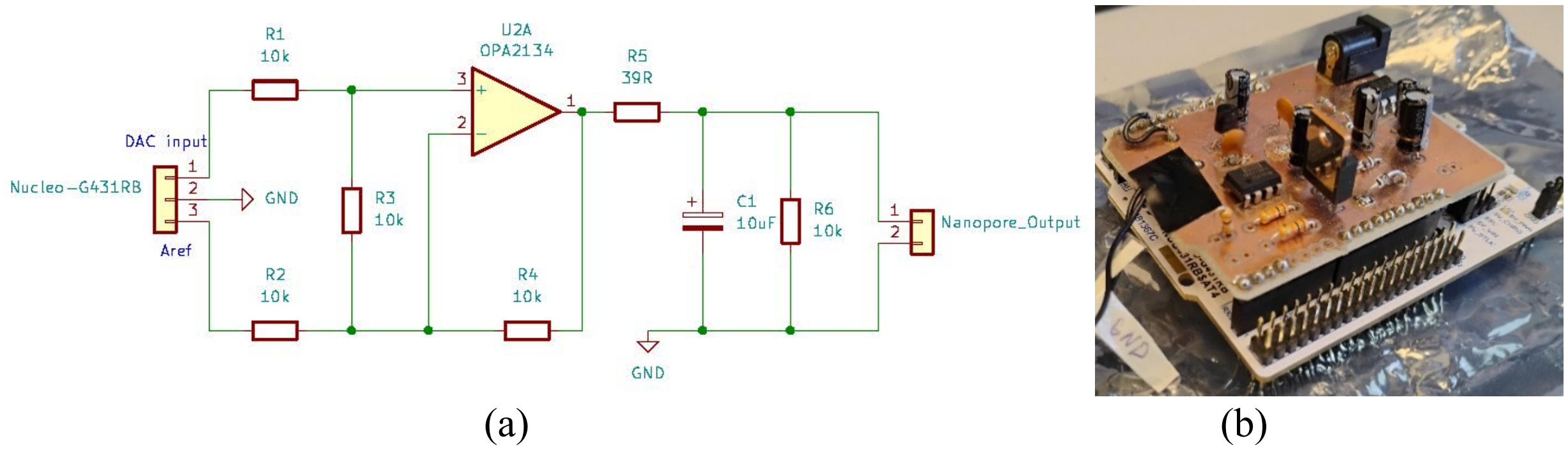

(a) (b)

*Figure 2. Self-made programmable voltage controller for nanopore-SERS measurements. (a) Core circuit diagram of the bipolar voltage-output stage. (b) Physical implementation of the compact custom circuit.*

The PCB includes an NE555-based negative-voltage generator adapted from a standard charge-pump design[2], regulated positive and negative supply rails, and an OPA2134-based analog output stage. A simplified schematic of the analog output stage is shown in Figure 2(b). Voltage commands are sent from the Python control program to the Nucleo board through USB serial communication. After receiving a command, the microcontroller writes the corresponding DAC code, and the PCB converts the resulting unipolar DAC signal into the required bipolar output voltage.

This design allows the software to reverse the voltage polarity or adjust the voltage magnitude during Raman measurements without manually reconnecting the electrodes. The programmable voltage range was selected to support both nanoparticle trapping and particle removal while avoiding unnecessarily high bias at the electrodes. When the Raman signal indicates an unwanted particle-associated state, the applied voltage can therefore be reversed or adjusted directly by the control software.

### 2.3 Peak detection pipeline for real-time SERS analysis

Real-time feedback requires each SERS spectrum to be analyzed before the next spectrum arrives, despite baseline drift and noise fluctuations. To meet these requirements, we used a five-stage pipeline to detect candidate peaks, refine their parameters, and assign an operational state to each spectrum. For the spectrum acquired at frame $t$, this state is denoted by s_t and is used for voltage control. The five stages are described below.

**A. Spectral preprocessing**

Each acquired spectrum was first corrected for cosmic-ray spikes using a modified Z-score approach adapted from Whitaker and Hayes[3]. This correction was applied before candidate-region detection because isolated, high-amplitude cosmic-ray spikes could otherwise satisfy the signal threshold and propagate as false peak candidates into the subsequent continuous wavelet transform (CWT) analysis. The selected Raman-shift window was then interpolated onto a uniform axis. Uniform sampling was required because the physical interpretation of the CWT scale depends on the spacing between adjacent Raman-shift points, thereby providing a consistent basis for wavelet-based position and width estimation.

**B. Candidate peak region detection**

For frame $t$, let $x_i$ denote the Raman shift at uniformly sampled point $i$, and let $y_{\{t,i\}}$ denote the corresponding preprocessed intensity. A running local-median curve was calculated to estimate the slowly varying spectral baseline. The local baseline $b_{\{t,i\}}$ was defined as

$$b_{t,i} = \text{median}\{y_{t,j}: j \in W_{t,i}\},$$

where $W_{t,i}$ is the set of point indices in the local window centered at $x_i$, and $j$ indexes the points within this window. The baseline was estimated locally to accommodate the slowly varying background across the Raman-shift range. A median filter was used because it is less strongly displaced by high-intensity Raman features than a local mean, providing a more stable reference for candidate detection. The residual intensity at point $i$ in frame $t$ was calculated as $r_{t,i} = y_{t,i} - b_{t,i}$. A spectrum-level robust noise estimate based on the median absolute deviation (MAD), denoted by $\sigma_{\{\text{MAD},t\}}$ , was then calculated from the residuals as

$$\sigma_{\text{MAD,t}} = 1.4826 * \text{median}_{\text{t}}\left|r_{t,i} - \text{median}(r_t)\right|.$$

Here, $r_t = r_{t,i}$ denotes the residual spectrum in frame t, and $\text{median}_{\text{I}}$ denotes the median over all analyzed point indices $i$. The factor 1.4826 makes the MAD consistent with the standard deviation for Gaussian noise. MAD was used because variance-based noise estimates can be disproportionately increased by strong spectral features and occasional high-amplitude fluctuations; it therefore characterizes the typical residual variation without allowing a small number of extreme points to dominate the detection threshold. A binary candidate indicator $m_{\{t,i\}}$ was then assigned to each point according to

$$m_{t,i} = \begin{cases} 1, y_{t,i} \geq b_{t,i} + 3\sigma_{MAD,t}, \\ 0, y_{t,i} < b_{t,i} + 3\sigma_{MAD,t}. \end{cases}$$

Thus, $m_{t,i} = 1$ indicates that the intensity exceeds the local baseline by at least three robust noise standard deviations, whereas $m_{t,i} = 0$ indicates that it does not. Combining the position-dependent baseline with a spectrum-level noise estimate allows the criterion to respond to both baseline variation across Raman shifts and differences in noise amplitude between acquired spectra. Let $M_{t,l}$ denote the $l$-th contiguous set of point indices for which $m_{t,i} = 1$ in frame $t$, and let $M_t = M_{t,l}$ denote the set of all retained candidate regions in that frame. A region was retained only when $\left|M_{t,l}\right| > 2$, where $\left|M_{t,l}\right|$ is the number of sampled points in the region. This requirement rejects isolated one- or two-point excursions that are less consistent with a spectrally resolved Raman band. The screening stage also acts as a computational gate: spectra for which $M_t = \varnothing$ bypass CWT analysis and peak fitting, whereas retained regions constrain the subsequent ridge construction and fitting.

**C. CWT-based peak localization**

Spectra containing candidate regions were further analyzed using CWT-based ridge detection. Let $y_{t(x)}$ denote the interpolated intensity of frame $t$ as a function of Raman shift $x$. Its CWT coefficient $C_t(a, x_0)$ at wavelet scale $a$ and Raman-shift position $x_0$ was calculated as

$$C_t(a, x_0) = \frac{1}{\sqrt{a}} \int y_t(x) \psi\left(\frac{x - x_0}{a}\right) dx,$$

where $\psi$ denotes the second-derivative Gaussian analyzing wavelet. The multiscale representation was used because SERS peaks can vary in width and can overlap with neighboring spectral structures. A structured peak is expected to produce positionally related maxima over a range of wavelet scales, whereas maxima arising from local noise are generally less persistent across scales. Local maxima in $C_t(a, x_0)$ were therefore identified at individual scales and linked across scales to form ridge structures [4]. Ridge seeding and grouping were constrained to the candidate regions in $M_t$, and only ridges extending across a sufficient range of scales were retained.

Compared with conventional CWT peak picking, we used the small-scale portion of each accepted ridge specifically for fitting initialization. The ridge endpoint at the smallest analyzed scale defined a local trajectory segment, within which the largest wavelet coefficient and its corresponding scale provided the initial peak center and width, respectively. This design was intended to retain narrow components adjacent to broader structures; the resulting estimates served only as initial values, whereas the final peak centers and widths were determined by pseudo-Voigt fitting.

**D. Peak parameter refinement**

CWT provides approximate peak locations and width-related scale information, whereas spectral-state determination requires quantitative full-width-at-half-maximum (FWHM) estimates. The ridge-derived parameters were therefore used to initialize nonlinear pseudo-Voigt fitting rather than being treated as the final peak measurements [5]. For an individual peak, the fitted profile $P(x)$ at Raman shift $x$ was defined as

$$P(x) = A[\eta L(x; x_c, \Gamma) + (1 - \eta)G(x; x_c, \Gamma)] + c,$$

where $L(x; x_c, \Gamma)$ and $G(x; x_c, \Gamma)$ denote unit-amplitude Lorentzian and Gaussian components, respectively, with peak center $x_c$ and FWHM $\Gamma$. The parameter $A$ is the peak amplitude, $\eta \in [0,1]$ is the Lorentzian-Gaussian mixing fraction, and $c$ is a constant local baseline offset. The adjustable mixing fraction allows the model to represent line shapes intermediate between the Gaussian and Lorentzian limits, while multiple ridge-initialized components can be fitted within the same candidate region to represent partially overlapping features. The peak-center estimate obtained from the small-scale endpoint of each accepted ridge, together with its ridge-derived width estimate, was used to initialize the fit. The set of successfully fitted peaks in frame $t$ was represented as $F_t = \{p_{t,k}\}_{k=1}^{K_t}$, where $K_t$ is the number of fitted peaks and $p_{t,k}$ is the $k$-th peak, characterized by center $x_{c,t,k}$, FWHM $\Gamma_{t,k}$ amplitude $A_{t,k}$, mixing fraction $\eta_{\mathrm{t,k}}$, and baseline offset $c_{\mathrm{t,k}}$.

**E. Spectral-state determination.**

The peak-detection results were converted into the operational spectral state $s_t$ for feedback control. This step was not intended to infer molecular identity or directly count particles, but to reduce the extracted spectral features to categories that could be evaluated by the voltage controller. The presence of at least one retained region in $M_t$ was used to determine whether a detectable candidate signal was present. Let $\Omega$ denote the set of point indices in the analyzed Raman-shift range and let $|\Omega|$ denote the number of points in that set. The mean local-median baseline in frame $t$ was calculated as $\bar{b}_t = |\,\Omega\,|^{-1} \sum_{i\in\Omega} b_{t,i}$ . For spectra with successfully fitted peaks, the maximum fitted FWHM was defined as $\Gamma_t^{\max} = \max_{1\le k\le K_t} \Gamma_{t,k}$ . The prescribed classification thresholds $T_b$ and $\Gamma_{\max}$ denote the upper limits on the mean baseline and maximum fitted FWHM, respectively, for a single-particle-like state.

Finally, the operational spectral state was assigned as follows:

$$s_t = \begin{cases} \text{none}, & M_t = \varnothing, \\ \text{single}, & M_t \neq \varnothing, \bar{b}_t < T_b, \Gamma_t^{\max} < \Gamma_{\max}, F_t \neq \varnothing, \\ \text{multi}, & \text{otherwise}. \end{cases} \quad (1)$$

The labels are control-oriented spectral categories: none denotes the absence of a retained candidate region,single denotes a candidate signal with a successfully fitted peak set, a mean baseline below $T_b$, and a maximum fitted FWHM below $\Gamma_{max}$, andmulti denotes all remaining signal-containing conditions, including elevated-background, broad-peak, and fitting-failure cases. These labels do not constitute independent measurements of particle number.

### 2.4 SERS-driven voltage control strategy

The applied voltage $V$ was adjusted in real time according to the spectral state $s_t$ determined by Eq. (1), and the control strategy is summarized in Algorithm 1. Here, $t$ is the frame index, $t_{max}$ is the total number of processed frames, and $\Delta t$ denotes the actual interval between consecutively acquired spectra. Raman spectra were captured at a nominal frequency of 10 Hz in the measurements reported here. The actual capture frequency was slightly lower because each acquisition cycle also included camera readout and data-transfer overhead. Although Algorithm 1 represents $\Delta t$ as constant for simplicity, the implemented software calculated $d_{\text{single}}$ and $d_{\text{hold}}$ directly from the recorded wall-clock timestamps. The variables $d_{\text{single}}$, $n_{\text{multi}}$, and $d_{\text{hold}}$ represent the duration of a continuous single-particle-like state, the number of consecutive multiparticle/high-background states, and the duration of a continuous no-signal state under negative voltage, respectively.

**Algorithm 1:** Spectral-state-driven voltage control

**Input :** Spectral states $s_t \in \{\text{none}, \text{single}, \text{multi}\}$; frame period $\Delta t$

**Output:** Applied voltage $V$

$V \leftarrow V_0$

$d_{\text{single}}, n_{\text{multi}}, d_{\text{hold}} \leftarrow 0, 0, 0$

**while** $t < t_{\max}$ **do**

- **if** $s_t = \text{none}$ **then**
  - $d_{\text{single}}, n_{\text{multi}} \leftarrow 0, 0$
- **if** $V > 0$ **then**
  - **switch** $s_t$ **do**
    - **case** single **do**
      - $d_{\text{single}} \leftarrow d_{\text{single}} + \Delta t$
      - $n_{\text{multi}} \leftarrow 0$
      - **if** $d_{\text{single}} \geq T_{\text{single}}$ **then**
        - $V \leftarrow -V$
        - $d_{\text{single}}, d_{\text{hold}} \leftarrow 0, 0$
    - **case** multi **do**
      - $n_{\text{multi}} \leftarrow n_{\text{multi}} + 1$
      - $d_{\text{single}} \leftarrow 0$
      - **if** $n_{\text{multi}} \geq N_{\text{multi}}$ **then**
        - $V \leftarrow -V$
        - $n_{\text{multi}}, d_{\text{hold}} \leftarrow 0, 0$
- **else**
  - **if** $s_t \neq \text{none}$ **then**
    - $d_{\text{hold}} \leftarrow 0$
    - **if** $V > V_{\min}$ **then**
      - $V \leftarrow V - \Delta V$
  - **else**
    - $d_{\text{hold}} \leftarrow d_{\text{hold}} + \Delta t$
    - **if** $d_{\text{hold}} \geq T_{\text{hold}}$ **then**
      - $V \leftarrow |V|$
      - $d_{\text{hold}} \leftarrow 0$
- Apply $V$ during acquisition of frame $t + 1$
- $t \leftarrow t + 1$

The system was initialized at $V_0 = +600$ mV, where $V_0$ denotes the initial applied voltage. When a single-particle-like signal persisted for $T_{\text{single}} = 4$ s, the voltage polarity was reversed to release the particle. Similarly, the voltage was reversed when a multiparticle/high-background signal persisted for $N_{\text{multi}} = 5$ consecutive frames. During the negative-voltage phase, if a signal remained detectable, the voltage was decreased in steps of $\Delta V = 100$ mV to a lower limit of $V_{\min} = -800$ mV. Once the no-signal state persisted for $T_{\text{hold}} = 7$ s, the voltage was returned to a positive value and the system re-entered the particle-capture phase.

## 3 Experimental results and discussion

### 3.1 Raman Peak Extraction and Spectral-state Classification

To evaluate whether the algorithm could extract features suitable for real-time feedback, we analyzed representative single-particle-like and multiparticle/high-background spectra.

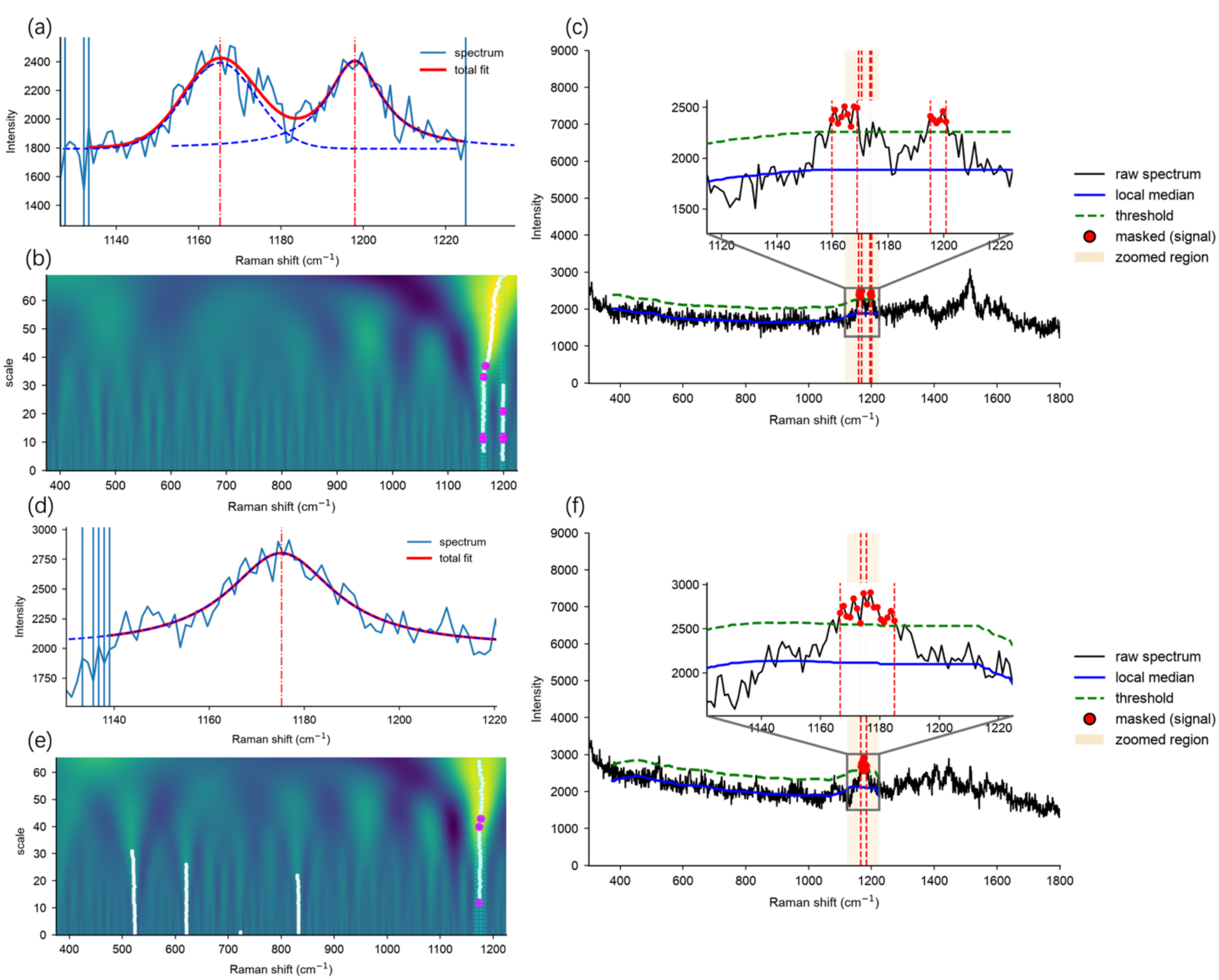

*Figure 3. Representative peak-detection results for single-particle-like and multiparticle/high-background SERS spectra. (a, d) Pseudo-Voigt fitting of the candidate peak regions for the single-particle-like and multiparticle/high-background spectra, respectively. (b, e) Corresponding CWT coefficient maps with detected ridge trajectories. (c, f) Full Raman spectra and enlarged candidate regions showing the local median, adaptive threshold, and signal mask.*

As shown in Figure 3, candidate regions were identified using the local median and an MAD-based adaptive threshold. CWT ridge detection and pseudo-Voigt fitting were then applied to extract peak positions, FWHMs, and baseline levels. The representative single-particle-like spectrum in Figure 3. (a-c) had a mean baseline of 1759.71 counts and a maximum FWHM of 20.74 $cm^{-1}$. Both values satisfied the classification thresholds of $T_b = 2000$ counts and $\Gamma_{max} = 25$ $cm^{-1}$. In contrast, the spectrum in Figure 3(d-f) had a mean baseline of 2090.27 counts and a maximum FWHM of 27.24 $cm^{-1}$, and was therefore classified as multiparticle/high-background. The number of Raman peaks was not used as an indicator of particle number.

Among the 61 spectra containing candidate signals, 25 were classified as single-particle-like and 36 as multiparticle/high-background (see Table 1). The median baseline was 1087.73 and 3029.58 counts, respectively, while the median maximum FWHM was 11.85 and 16.99 $cm^{-1}$, respectively. Peak-fitting failures were identified from exceptions recorded during real-time analysis. Fitting failed in 4 of the 61 spectra, corresponding to a failure rate of $4/61 = 6.6\%$.

The median processing times were 29.41 ms for single-particle-like spectra and 27.20 ms for multiparticle/high-background spectra. These processing times were shorter than the nominal 100 ms capture period corresponding to a capture frequency of 10 Hz. Because each acquisition cycle also included camera readout and data-transfer overhead, the interval between consecutively delivered spectra was slightly longer than 100~ms. The processing results therefore support real-time operation under the acquisition conditions used here. These labels represent operational spectral states used for feedback control rather than independently verified particle counts.

### 3.2 Real-time spectral-state-guided voltage control

Figure 4 shows the synchronized Raman spectra, algorithm-assigned spectral states, and applied voltage during real-time operation. The system initially applied +600 mV to capture particles. Several transient single-particle-like states were detected during the measurement, but they did not persist for the required 4 s and therefore did not trigger voltage reversal. Likewise, isolated multiparticle/high-background states that did not meet the consecutive-frame criterion caused no voltage change.

Frames 477-492 provide a representative control event. Frames 477-482 were classified as single-particle-like, but their duration was only approximately 0.6 s, so the voltage remained at +600 mV. Frames 485-490 were subsequently classified as multiparticle/high-background and satisfied the trigger condition, causing the voltage to reverse from +600 to -600 mV. Because a signal remained detectable in frame 491, the voltage was further decreased to -700 mV. No signal was detected from frame 492 onward. After this no-signal state persisted, the voltage was restored to +600 mV following frame 628, allowing the system to resume particle capture.

These results show that the system rejected transient state fluctuations and adjusted the voltage only when the predefined duration or consecutive-frame criteria were met. If a Raman signal persisted after polarity reversal, the reverse-bias magnitude was increased in 100 mV steps. The disappearance of the Raman signal indicates spectral clearance from the observation region, but does not independently confirm complete particle removal from the nanopore.

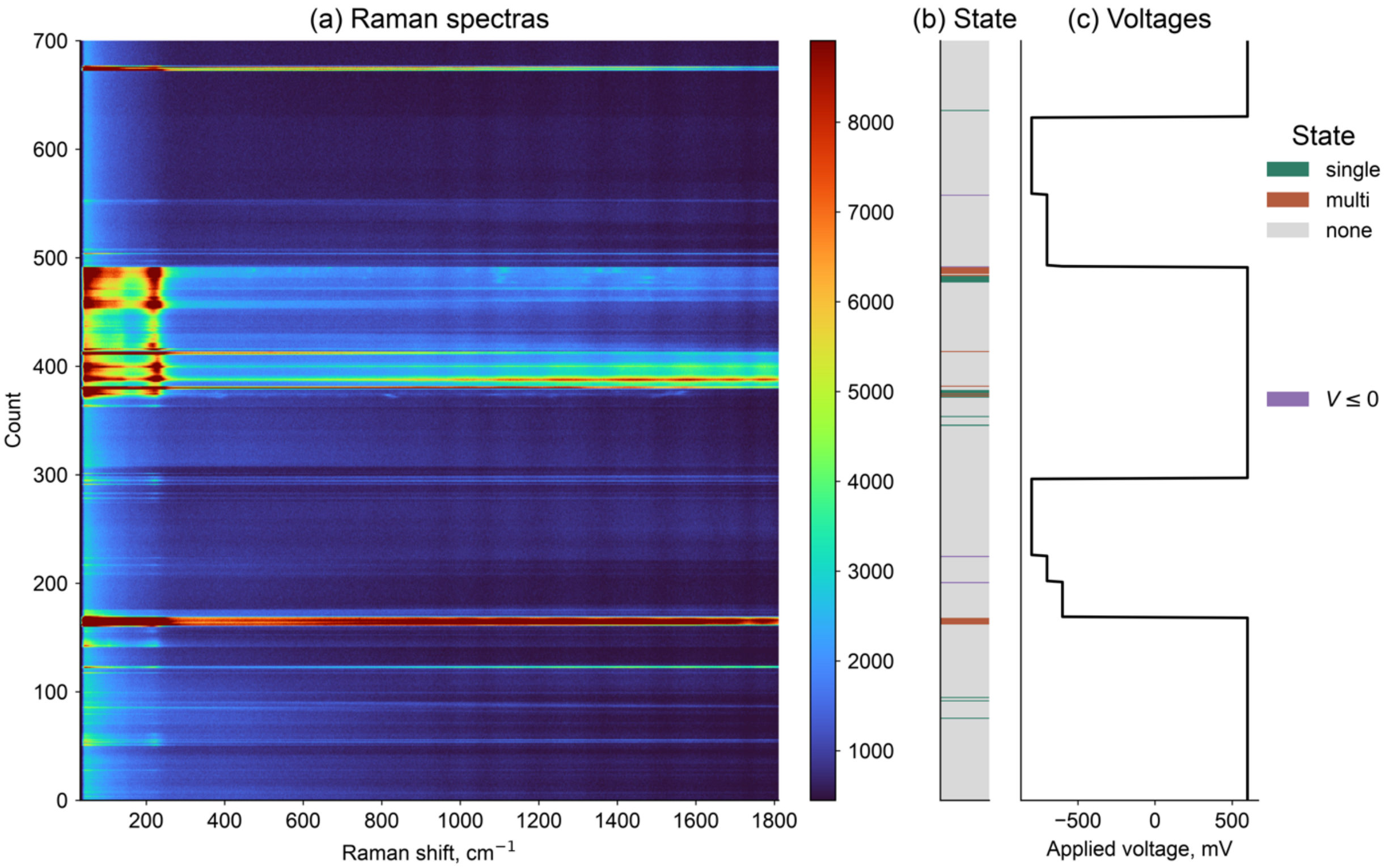

*Figure 4. Representative closed-loop operation of the SERS-driven voltage-control system. (a) Raman-intensity map as a function of Raman shift and acquisition frame. (b) Algorithm-assigned spectral state for each frame; purple marks indicate frames acquired at $V \leq 0$. (c) Applied voltage aligned with the spectra and states by acquisition frame. The voltage polarity was reversed when a state-persistence criterion was met, and the reverse-bias magnitude was increased when a particle-associated signal remained detectable.*

### 3.3 Discussion

Taken together, these results support the feasibility of using spectral features to control the applied voltage within the acquisition cycle. Peak detection and state classification required approximately 27-29 ms, leaving substantial time within the nominal 100 ms capture period. With this computational margin, the controller could base voltage changes on persistent states rather than react to every classified frame. Requiring a state to persist suppressed responses to short-lived spectral fluctuations but also delayed intervention after a genuine transition. The appropriate balance between stability and response speed will therefore depend on the trapping dynamics and the intended measurement.

The single-particle-like spectra showed a lower median baseline (1087.73 counts) and a narrower median maximum FWHM (11.85 $cm^{-1}$), whereas the multiparticle/high-background spectra exhibited a higher median baseline (3029.58 counts) and broader fitted features (16.99 $cm^{-1}$). This concurrent increase in baseline level and apparent peak width is consistent with greater spectral complexity, potentially arising from multiple enhancement sites, heterogeneous local environments, or partially overlapping spectral components. However, because particle occupancy was not independently measured, these spectral characteristics cannot be directly interpreted as particle-number signatures.

The physical interpretation of the control response nevertheless remains limited by the available measurements. Loss of the Raman signal after reverse bias confirms that detectable SERS enhancement disappeared but does not distinguish particle ejection from displacement outside the hot spot. Moreover, the detection thresholds, persistence criteria, voltage increment, and minimum reverse voltage were selected during initial trials and were not systematically optimized. The present results therefore demonstrate operation of the closed-loop workflow rather than universal classification or control settings. Repeated measurements across devices, supported by independently labeled particle states, will be needed to determine robust operating ranges.

## 4 Conclusion

We described the microcontroller system which allows to remove clogged particles from the nanopore automatically. This system controls voltage based on collected Raman spectra. The applied voltage information is stored together with the acquired Raman spectra. These results demonstrate that the proposed algorithm enables automated extraction of Raman signal parameters and classification of spectra into several categories: no detectable signal, signal from a single particle, signal originating from aggregated particles. This system can be used for further analysis which is necessary to create scientific equipment for fast sequencing of low concentrated molecules in real time.

**Reference**

[1]. Huang J A, Mousavi M Z, Zhao Y, et al. SERS discrimination of single DNA bases in single oligonucleotides by electro-plasmonic trapping[J]. Nature Communications, 2019, 10(1): 5321.

[2]. R. Mitchell, “Build Your Own Negative Voltage Generator,” All About Circuits, Jun. 7, 2018. [Online]. Available: https://www.allaboutcircuits.com/projects/build-your-own-negative-voltage-generator/. Accessed: [05-09-2026].

[3]. Whitaker D A, Hayes K. A simple algorithm for despiking Raman spectra[J]. Chemometrics and Intelligent Laboratory Systems, 2018, 179: 82-84.

[4]. Du P, Kibbe W A, Lin S M. Improved peak detection in mass spectrum by incorporating continuous wavelet transform-based pattern matching[J]. bioinformatics, 2006, 22(17): 2059-2065.

[5]. Lange E, Gröpl C, Reinert K, et al. High-accuracy peak picking of proteomics data using wavelet techniques[M]//Biocomputing 2006. 2006: 243-254.